# Implication of Kinetic Alfvèn Waves to Magnetic Field Turbulence Spectra: Earth's Magnetosheath

N. K. Dwivedi[*,1], S. Kumar[2,3], P. Kovacs[4], E. Yordanova[5], M. Echim[6], R. P. Sharma[7], M. L. Khodachenko[1,8,9], Y. Sasunov[1,8]

[1]*Space Research Institute, Austrian Academy of Sciences, Schmiedlstraß 6, 8042 Graz, Austria*

[2]*School of Space Research, Kyung Hee University, Yongin, Gyeonggi-Do, 446-701, Korea*

[3]*Shandong Provincial Key Laboratory of Optical Astronomy and Solar-terrestrial Environment, Institute of Space Science, Shandong University, Weihai, China*

[4]*Mining and Geological Survey of Hungary, Budapest, Hungary*

[5]*Swedish Institute of Space Physics, Uppsala, Sweden*

[6]*Institut Royale d'Aeronomie Spatiale de Belgique, 1180 Brussels, Belgium*

[7]*Centre for Energy Studies, Indian Institute of Technology Delhi, India*

[8]*Skobeltsyn Institute of Nuclear Physics, Moscow State University, Moscow, Russia*

[9]*Institute of Astronomy, Russian Academy of Science, Moscow 119017, Russian Federation*

## Abstract

In the present paper, we investigate the power-law behaviour of the magnetic field spectra in the Earth's magnetosheath region using Cluster spacecraft data under solar minimum condition. The power spectral density of the magnetic field data and spectral slopes at various frequencies are analysed. Propagation angle, $\theta_{kB}$, and compressibility, $R_\parallel$, are used to test the nature of turbulent fluctuations. The magnetic field spectra have the spectral slopes, $\alpha$, between -1.5 to 0 down to spatial scales of $20\rho_i$ (where $\rho_i$ is ion gyroradius), and show clear evidence of a transition to steeper spectra for small scales with a second power-law, having $\alpha$ between -2.6 to -1.8. At low frequencies, $f_{sc} < 0.3 f_{ci}$ (where $f_{ci}$ is ion gyro-frequency), $\theta_{kB}$~ 90° to the mean magnetic field, $B_0$, and $R_\parallel$ shows a broad distribution, $0.1 \leq R_\parallel \leq 0.9$. On the other hand at $f_{sc} > 10 f_{ci}$, $\theta_{kB}$ exhibits a broad range, $30° \leq \theta_{kB} \leq 90°$, while $R_\parallel$ has a small variation: $0.2 \leq R_\parallel \leq 0.5$. We conjecture that at high frequencies, the perpendicularly propagating Alfvén waves could partly explain the statistical analysis of spectra. To support our prediction of kinetic Alfvèn wave dominated spectral slope behaviour at high frequency, we also present a theoretical model and simulate the magnetic field turbulence spectra due to nonlinear evolution of kinetic Alfvèn waves. The present study also shows the analogy between the observational and simulated spectra.

*Corresponding author: N. K. Dwivedi

*Email: navin.dwivedi@gmail.com

# 1 Introduction

Plasma turbulence ensues diversely in the laboratory, space, and astrophysical plasmas under a variety of conditions (Bruno & Carbone 2005). Turbulent plasmas display complex, chaotic, broadband fluctuations, which are interpreted as a scale-invariant cascade of energy from large scales, where the energy is injected, to small-scale where the energy is dissipated. Turbulence is important in many aspects. It influences the particle heating and acceleration, governs the dynamics of the Earth's magnetosphere, as well as energy and mass inflow from the solar wind to the magnetosphere. The transition of solar wind plasmas through the bow shock manifests non-adiabatic changes in the ion distribution (Gosling et al. 1989), the significant processing of these distributions within the magnetosheath and in the boundary layers of the magnetosphere is still a matter of extensive investigation and research. Study of magnetic field fluctuations in the Earth's magnetosheath is of great importance to understand space plasma phenomena like turbulence, particle acceleration in the boundary regions associated with the Earth's bow shock waves and magnetopause. One of the pivotal problems in space and planetary science is how the solar wind influences the plasma processes inside the bow shock, particularly in the magnetosheath. This research topic is not only vital for the terrestrial planets like Earth but also for the planets without intrinsic magnetic fields such as Venus and Mars.

The solar wind is an example of freely developed turbulence, the Earth's magnetosheath, on the contrary, is a domain bounded by the bow shock and the magnetopause, where an important energy injection takes place at the ion scales. As compared to solar wind turbulence, less attention is paid to investigate the plasma turbulence in planetary environments such as magnetopause, magnetotail and magnetosheath regions of planetary bodies. So far the plasma turbulence studies are limited to the planetary bodies like Mercury (Uritsky et al. 2011), Venus (Vörös et al. 2008; Dwivedi et al., 2015), Earth (Zimbardo et al. 2010), Jupiter (Tao et al., 2015), and Saturn (Bavassano Cattaneo et al. 2000; von Papen et al. 2013; Hadid et al. 2015). In case of Saturn's magnetosheath, Hadid et al. (2015) detected that the spectral index ranges from -1.2 at low frequencies down to -2.5 at high frequencies (higher than the proton gyrofrequency) and did not observe the inertial range Kolmogorov-type spectra which usually are observed in the case of solar wind. The absence of Kolmogorov-type scaling in the Earth's magnetosheath has also been shown by Zimbardo et al. (2010) and Tao et al. (2015). In the presence of boundaries, such as the bow shock, the turbulent fluctuations are processed/compressed through the bow shock, and the instabilities induced by the shock can have a larger impact on the pre-existing turbulence, destroying the pre-existing fluctuations and their nonlinear relationships. While the aforementioned signature of the turbulence in the magnetosheath is observed, its interpretation is complicated due to the presence of different characteristic kinetic spatial and time scales. The interpretation is further complicated by the fact that the measured spectrum of plasma fluctuations consists of both the fluctuations in the plasma frame and the spatial structures that are swept by the spacecraft. A statistical study on turbulence at subproton-and-electron-scale in the terrestrial magnetosheath using the Cluster/STAFF search coil magnetometer data demonstrated clear spectral breaks near the electron scale separating two power-law like frequency bands namely dispersive and dissipation ranges (Huang et al. 2014). Huang et al. (2017) have shown that the turbulence properties at MHD scale, as is well known in the solar wind, do not persist after interaction with the Earth's bow shock, and random-like fluctuations were generated behind the bow shock, having the $f^{-1}$ spectrum in the low-frequency range. Later, the electric and magnetic field spectra have been simultaneously presented in the Earth's magnetosheath, having same behaviour in the MHD regime, however, showing different behaviour in the kinetic range (Matteini et al. 2016). In a recent study, the in situ measurement of Magnetospheric MultiScale spacecraft were used to provide new insights into the nature of turbulence in the Earth's magnetosheath (Breuillard et al. 2018).

In this paper, we first discuss the in situ measurement of magnetic field turbulence spectra and associated spectral indices in terms of ion characteristic space and temporal scales in the magnetosheath region of the Earth, and secondly, we present a theoretical aspect of kinetic Alfvèn waves (KAWs) driven turbulence spectra and spectral indices. Finally, we compare our simulated magnetic field spectra and spectral indices with the observational spectra. The present statistical investigation and the obtained results show that in the majority of the cases, the magnetic field spectra do not exhibit the inertial range at the low frequencies. However, at high frequencies, the spectral index varies between -2.6 to -1.8 and hence support the claim made by Zimbardo et al. (2010) and Tao et al. ( 2015). Our theoretical model provides the physical interpretation of the observed spectra and related spectral indices and claims that the nonlinear KAWs could be one of the possible reasons for the steepening in the spectra at higher frequencies (or at small scale). The paper is structured in the following sections: Section 2 describes the data selection procedure and a method adopted for computing the magnetic field spectra and spectral slopes. Section 3 presents the two-fluid model, the intermediate steps of the derivation part of the dynamical equation of KAWs, and a brief description of the adopted numerical simulation approach. Section 4 presents the results obtained by the Cluster observation and comparisons with the results obtained by numerical simulations. Section 5 gives the physical insights on the nonlinear processes leading to the turbulence in the magnetosheath region. Section 6 concludes on the obtained results.

## 2. Cluster observation

### 2.1 Data Selection

We analyse Cluster-1 magnetometer (Balogh et al. 2001), and the Cluster Ion Spectroscopy (CIS) experiment (Remé et al. 2001) measurements in the terrestrial magnetosheath in the years 2007-2008. The *22.5 Hz* high-resolution Cluster-1 fluxgate magnetometer data is used and *337* magnetosheath events with a fixed duration of *35* minutes are selected. The magnetosheath events selection is based on the simultaneous scanning of the following parameters: spacecraft position, magnetic field magnitude, ion velocity, ion temperature and omnidirectional ion energy flux. We apply the following ranges for the magnetosheath events selection: day-side spacecraft positions ($X_{GSE} > 0$), magnetic field magnitude < *70 nT*, ion bulk velocity< *450 kms$^{-1}$*, ion temperature > 0.5 MK, and ion energy flux > *0.2 keVcm$^{-2}$s$^{-1}$sr$^{-1}$* in the energy range *0.1-12 keV*. To exclude the mixing of ion foreshock region with the magnetosheath region while making the events selection, we have rejected those time intervals in which the count rate of the highest energy channels (*16-30 keV*) exceeded the threshold of the ion energy flux, i.e., 0.33 *keVcm$^{-2}$s$^{-1}$sr$^{-1}$*. The electron foreshock is excluded by rejecting the intervals in which the wave energy measured by WHISPER (Waves of High frequency and Sounder for Probing of the Electron density by Relaxation) (Décréau et al. 1997) instrument observe the electric field spectra in the range of *2-82 kHz*. In this range, the electrostatic and electromagnetic natural emissions close to plasma frequency are detected. The mixing of the electron foreshock region with the magnetosheath region is also avoided by not considering those periods where the wave energy is enhanced due to Langmuir waves typical for this region. The applied ranges in the selection procedure also assume different values due to a month-scale variability in the parameters. The selection criteria work only when the data of all parameters are simultaneously available.

**2.2 Magnetic field spectra**

The selected magnetic field data in the magnetosheath are transformed into a Mean-Field-Aligned (MFA) coordinate system whose *z*-axis is parallel to the large-scale mean magnetic field direction; the MFA *x*-axis is perpendicular to the plane defined by the mean magnetic field and the spacecraft position; the MFA *y*-axis completes the right-handed coordinate system. The background or large-scale magnetic field is estimated from the average over the total duration of each time-series. The Welch algorithm (Welch 1967) is applied for the estimation of the power spectral density. Since the Welch method can result in erroneous spectral points at small frequencies, the first three data pairs of the Welch spectra are skipped. Furthermore, to reduce the instrumental noise at high frequencies, the spectra are truncated at frequencies where the signal to noise ratio falls below 2 (see the next section for further details).

The power spectral density is computed for the total magnetic field, $B_t$, in the spacecraft frequency frame, $f_{sc}$, normalised to ion gyro-frequencies, $f_{ci} = eB_0/2\pi m_i c$. Also, the power spectral density (PSD) of $B_t$ is estimated in the spatial spanning of streamline wave number in the spacecraft reference frame, $k = 2\pi f_{sc}/V_{flow}$, normalised to the inverse of the ion gyroradius $\rho_i = V_{T_i}/2\pi f_{ci}$. The values of $f_{ci}$ and $\rho_i$ are determined for each event selected in the magnetosheath using the onboard plasma measurements of the spacecraft. Note that for transforming the spacecraft frequencies to wave numbers, we suppose that the wave propagations in the rest frame are negligible in comparison to the speed of the bulk plasma flow, that is the Taylor's frozen-in flow hypothesis (Taylor 1938) is always applicable. Since the validity of the Taylor hypothesis is a-priori not known for a single event, we investigate the obtained statistical results in terms of different flow conditions. The conditional analyses prove that our results reliably characterise the spectral behaviour of the MS magnetic time-series (for details, see the Discussion part).

The spectral behaviour of magnetic field fluctuations is investigated with an automatic data analysis algorithm that identifies frequency ranges characterised by the same power-law; the algorithm also computes the corresponding spectral slope, ($\alpha$), for each identified frequency range. First, the raw PSD is computed with the (Welch 1967) algorithm and is subsequently re-sampled in frequency bins of equal logarithmic size. Then we define a sliding window of six points (frequencies) and move it, one point at a time, across the re-sampled PSD. At each step, we compute the spectral index resulting from the fit of PSD values within the sliding window and assign it to the mean frequency of the window. Thus we create a statistical ensemble of spectral slopes, $\alpha(f)$. At the end, an $[f_1, f_2]$ range is considered as a scaling range if all $\alpha(f)$ within the range ($f_1 \leq f \leq f_2$) takes a value in the domain defined as $\alpha(f_1) - L \leq \alpha(f) \leq \alpha(f_1) + L$, where $L$ is a predefined limit. In the analysis $L$ was defined dynamically, as the three-tenth of the whole range of variation of $\alpha(f)$ obtained for a given time interval. The authors have already used the same numerical procedure in a previous publication (for detail see Dwivedi et al. 2015). The results are shown in two-dimensional colour-coded histograms (Figs. 3-4). We also plot the number of occurrences of different α values for the whole frequency range, as shown in the right panels (Figs. 3-4). The histogram evidences the most probable $\alpha$ values for different frequency ranges. To exemplify the role of the ion temporal (ion gyrofrequency) and spatial scales (ion gyroradius), we plot the normalised two-dimensional (2D) histograms showing the spectral ranges and their scaling exponents for spectra normalised with the above characteristic scales (see the next section for further details).

Additionally, we create two additional statistical datasets: (i) a statistical ensemble of $R_\parallel$, the compressive power (defined below) as a function of frequency, and (ii) a statistical ensemble of the relative orientation $\theta_{kB}$ of a minimum variance direction concerning the mean magnetic field. The minimum variance analysis (McPherron et al., 1972) consists in finding the direction of the minimum variance of the mean magnetic field. Analytically, this direction is calculated by solving for the eigenvector associated with the minimum eigenvalue of the spectral matrix, written in the MFA system as follows:

$$R_{MFA}(f) = \begin{bmatrix} \tilde{B}_{\perp 1}^2(f) & \tilde{B}_{\perp 1}(f)\tilde{B}_{\perp 2}(f) & \tilde{B}_{\parallel}(f)\tilde{B}_{\perp 1}(f) \\ \tilde{B}_{\perp 1}(f)\tilde{B}_{\perp 2}(f) & \tilde{B}_{\perp 2}^2(f) & \tilde{B}_{\parallel}(f)\tilde{B}_{\perp 2}(f) \\ \tilde{B}_{\perp 1}(f)\tilde{B}_{\parallel}(f) & \tilde{B}_{\perp 2}(f)\tilde{B}_{\parallel}(f) & \tilde{B}_{\parallel}^2(f) \end{bmatrix} \quad (1)$$

where $\tilde{B}_{\perp 1}(f)$, $\tilde{B}_{\perp 2}(f)$ and $\tilde{B}_{\parallel}(f)$ are the spectral/Fourier components of the magnetic field transformed in the MFA system. With the assumption that the spectrum of the fluctuations is due to a superposition of plane waves one can assign the minimum variance direction to the direction of propagation of the wave. We also estimate the angle ($\theta_{kB}$) between the minimum variance eigenvector and the mean magnetic field ($B_0$) at each frequency, $f$. When $\theta_{kB} \approx 0°$, the fluctuations/waves would propagate mainly along $B_0$, thus parallel, and when $\theta_{kB} \approx 90°$, it is a perpendicular propagation. In general, the spectral matrix, Eq. (1) is non-diagonal. We define the compressibility ($R_{\parallel}$), as the ratio of the compressive power ($P_{\parallel}$) to the total power ($P_{total} = P_{\parallel} + P_{\perp}$), where $P_{\parallel}$, and $P_{\perp}$ are determined from the covariance or spectral matrix, Eq. (1).

## 3. Theoretical model and simulation

In the present study, we consider a fully ionized, homogenous, collisionless plasma consisting of electrons and ions. The two-fluid model of plasma for the steady state has been developed to analyse the nonlinear evolution of KAWs. The governing equation of KAW is first derived and then numerically solved to realise the turbulence spectrum.

## 3.1 Kinetic Alfvèn wave

The dynamical equation of KAW propagating in the *x-z* plane is derived under the two-fluid approach. The ambient magnetic field, $B_0$, is along the *z*-axis, Maxwell's equations and drift approximation are used to obtain the governing equation of KAW (Shukla and Stenflo, 1999, 2000; Dwivedi et al., 2013; Dwivedi & Sharma 2013):

$$\frac{\partial^2 \tilde{B}_y}{\partial t^2} = \lambda_e^2 \frac{\partial^4 \tilde{B}_y}{\partial x^2 \partial t^2} - \rho_i^2 V_A^2 \frac{\partial^4 \tilde{B}_y}{\partial x^2 \partial z^2} + V_A^2 \left(1 - \frac{n_e}{n_0}\right) \frac{\partial^2 \tilde{B}_y}{\partial z^2}. \tag{2}$$

where $T_e$, $T_i$ are the electron and ion temperatures, $m_e$, $m_i$ are the electron and ion masses, $\lambda_e = \sqrt{c^2 m_e / 4\pi n_0 e^2}$ is the collisionless electron skin depth, $V_A = \sqrt{B_0^2 / 4\pi n_0 m_i}$ is the Alfvén velocity, $\rho_i = V_{T_i}/\omega_{ci}$ is the ion gyroradius, $V_{T_i} = \left(\{\gamma_e k T_e + \gamma_i k T_i\}/m_i\right)^{1/2}$ is the acoustic speed, the electrons, and ions are assumed to be isothermal, i.e. $\gamma_e = \gamma_i = 1$, and $\omega_{ci} = eB_0/cm_i$ is the angular ion gyrofrequency. The adiabatic responses to the density are considered in the present study which gets modified in the presence of nonlinearity arises as KAW propagates along magnetic field lines ($B_0 \hat{z}$) (Dwivedi et al. 2013; Dwivedi & Sharma 2013):

$$\frac{n_e}{n_0} = \varphi\left(\tilde{B}\tilde{B}^*\right), \tag{3}$$

where $\varphi\left(\tilde{B}\tilde{B}^*\right) = \xi\left[\left|B_y\right|^2\right]$, $\xi = \left[(1-\varsigma(1+\delta))/16\pi n_0 T\right]\left(V_A^2 k_{0z}^2/\omega_0^2\right)$, $\varsigma = \omega_0^2/\omega_{ci}^2$, $\omega_0$ is the frequency of the pump KAW, $\delta = m_e k_{0x}^2 / m_i k_{0z}^2$, and $k_{0x}(k_{0z})$ is the component of the wave vector perpendicular (parallel) to $B_0 \hat{z}$.

We consider a plane wave solution to Eq. (2) as

$$B_y = \tilde{B}_{y0}(x,z,t)e^{i(k_{0x}x+k_{0z}z-\omega_0 t)}. \tag{4}$$

The dynamical equation of KAW is obtained using Eqs. (2) and (4) with the approximations $\partial_x \tilde{B}_0 \gg k_{0x}\tilde{B}_{y0}$, and $\partial_z \tilde{B}_0 \ll k_{0z}\tilde{B}_{y0}$

$$-\frac{2i\omega_0}{V_A^2 k_{0z}^2}\frac{\partial \tilde{B}_{y0}}{\partial t} - \frac{2i}{k_{0z}}\frac{\partial \tilde{B}_{y0}}{\partial z} - 2ik_{0x}\rho_i^2 \frac{\partial \tilde{B}_{y0}}{\partial x} - \rho_i^2 \frac{\partial^2 \tilde{B}_{y0}}{\partial x^2}$$
$$-\frac{\rho_i^2 k_{0x}^2}{k_{0z}^2}\frac{\partial^2 \tilde{B}_{y0}}{\partial z^2} - \left(\frac{\delta n_s}{n_0}\right)\tilde{B}_{y0} = 0. \tag{5}$$

One can write the governing equation in the dimensionless form as

$$i\frac{\partial \tilde{B}_{y0}}{\partial t} + 2i\Gamma_1 \frac{\partial \tilde{B}_{y0}}{\partial x} + \frac{\partial^2 \tilde{B}_{y0}}{\partial x^2} + i\frac{\partial \tilde{B}_{y0}}{\partial z} + \Gamma_2 \frac{\partial^2 \tilde{B}_{y0}}{\partial z^2} + \left|\tilde{B}_{y0}\right|^2 \tilde{B}_{y0} = 0. \tag{6}$$

where the dimensionless parameters are, $\Gamma_1 = k_{0x}\rho_i$ and $\Gamma_2 = \frac{\rho_i^2 k_{0x}^2}{2}$. The normalising factors are $z_n = 2/k_{0z}$, $t_n = \left(2\omega_0/V_A^2 k_{0z}^2\right)$, $x_n = \rho_i$ and $B_n = 1/\sqrt{\xi}$. Equation (6) describes the envelope dynamics of a linearly-polarised weakly-nonlinear KAW. The nonlinearity results from the frequency shift produced by the density fluctuations that adiabatically follow the variations of the wave amplitude.

**3.2 Numerical Simulation:**

A two-dimensional pseudo-spectral (2DPS) numerical method with periodic spatial domain $(2\pi/\kappa_x) \times (2\pi/\kappa_z)$ and grid $128 \times 128$ is exploited to obtain the solution of Eq. (6). The initial condition for the simulation is (see Kumar et al. 2009; Dwivedi et al. 2012; Dwivedi et al. 2013; Dwivedi & Sharma 2013):

$$\tilde{B}_{y0}(x,z,0) = B_{y00}\left(1+0.1\cos(\kappa_x x)\right)\left(1+0.1\cos(\kappa_z z)\right) \tag{7}$$

where the amplitude of the homogenous KAW is $B_{y00} = 1$, and the inverse of the spatial scales of the initial disturbances are $\kappa_x = \kappa_z = 0.1$. All fields are represented in discrete Fourier series having the integral wave-vector components. A fully de-aliased 2DPS numerical technique is implemented for the space integration and modified Gazdag predictor-corrector scheme is used for the transient evolution. To check the numerical accuracy of our computational methodology, we first exploited it for the 2D cubic nonlinear Schrödinger (NLS) equation and compared the numerical results with the well-known results of NLS. This allows us to replicate the associated instabilities precisely and then we employed the same numerical technique for Eq. (6). The following parameters and assumptions are used in the numerical simulation:

(i) Observed basic plasma parameters:

$B_0 = 1.5 \times 10^{-4} G$, $n_0 = 10 cm^{-3}$, $T_e = 2.6 \times 10^5 K$, $T_i = 2 \times 10^6 K$, $m_e = 9.1 \times 10^{-28} gm$,

$m_i = 1.67 \times 10^{-24} gm$, $c = 3 \times 10^{10} cms^{-1}$,

(ii) Basic assumptions:

$\omega_0 = 0.01 \times \omega_{ci}$, $k_{0x}\rho_i = 0.01$,

(iii) The derived parameters using basic parameters and assumptions:

$V_{Te} = 1.99 \times 10^8 cms^{-1}$, $V_{Ti} = 1.29 \times 10^7 cms^{-1}$, $V_A = 1.04 \times 10^7 cms^{-1}$,

$\lambda_e = 1.683 \times 10^5 cm^{-1}$, $\omega_{ce} = 2.64 \times 10^3 rads^{-1}$, $\omega_{ci} = 1.44 rads^{-1}$,

$\omega_{pe} = 1.78 \times 10^5 rads^{-1}$, $\rho_i = 8.92 \times 10^6 cm$, $k_{0z} = 1.38 \times 10^{-9} cm^{-1}$.

(iv) The values of normalising factors are:

$x_n = \rho_i = 8.92 \times 10^6 cm$, $z_n = 1.45 \times 10^9 cm$, $B_n = 7.6 \times 10^{-4} G$, $t_n = 1.4 \times 10^2 s$.

# 4. Results

## 4.1 Observational results

Figures 1-2 show the normalised superposed magnetic field spectra of the 337 Earth magnetosheath time intervals. In Fig. 1, the spectra are represented as a function of frequency ($f_{sc}/f_{ci}$), however in Fig. 2, the spectra are shown in the wavenumber ($k$) domain. For the cases of wave number spectra, Taylor's frozen-in hypothesis [15] is applied to map the spacecraft frequencies, $f_{sc}$, to wave numbers ($k$) according to the relation $k = \dfrac{2\pi f_{sc}}{V_{flow}}$, where $V_{flow}$ stands for the bulk velocity of the plasma. The magnetic field fluctuations are normalised to the mean magnetic field ($B_0$) in each spectrum. Additionally, the spacecraft frequencies ($f_{sc}$) and wave numbers ($k$) are normalised respectively to the characteristic temporal and spatial scales of the ions, i.e., the ion gyrofrequency, $f_{ci}$, and the reciprocal of the ion gyroradius, $\rho_i = \dfrac{V_{T_i}}{2\pi f_{ci}}$. The power spectra ($P(f_{sc})$) computed in the spacecraft frequency frame are then transformed into the frequency domain as:

$$f_{sc} \to \frac{f_{sc}}{f_{ci}}$$

$$P(f_{sc}) \to P\left(\frac{f_{sc}}{f_{ci}}\right) = P(f_{sc}) f_{ci}$$

and in the wavenumber domain as:

$$k \to k\rho_i$$

$$P(k) = \frac{P(f_{sc}) V_{flow}}{2\pi} \to P(k\rho_i) = \frac{P(k)}{\rho_i}$$

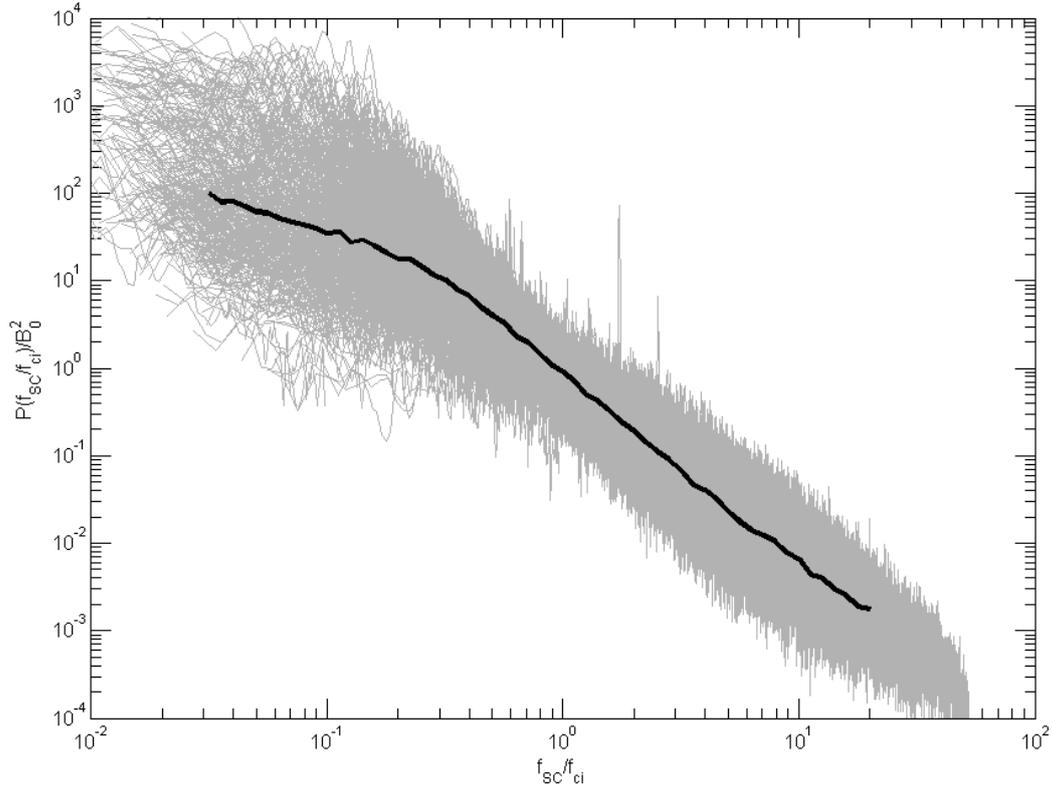

Figure 1. *Normalised superposed power spectral density of the magnetic field time-series (337 events) inside the magnetosheath. The frequency in the spacecraft frame, $f_{sc}$, normalised to the ion gyrofrequency $f_{ci}$ is shown on the x-axis, and the magnetic field power spectral density, normalised to the square of the mean magnetic field amplitude, $B_0^2$, is shown on the y-axis.*

The sensitivities of spacecraft fluxgate magnetometer instruments are limited and exhibit a noise level of $10^{-4} nT^2 Hz^{-1}$ at and beyond 1 *Hz*, for Cluster (Robert et al. 2014, Carr et al. 2014). Therefore, the investigated spectra are truncated at frequencies where the signal's energy densities are smaller than $2 \times 10^{-4} nT^2 Hz^{-1}$, i.e., the signal to noise ratio becomes less than 2. The grey backgrounds in the superposed spectra shown in Figs. 1-2 represent the individual spectra while the black curves show their medians. The mean trends of the spectra exhibit power-law behaviour in both domains (frequency and wave number), at low frequencies (small wave numbers) the spectra are slightly flatter and then followed by a steeper power-law trend at high frequencies. The scaling regions, the spectral slopes and the breaks of the spectra are statistically investigated with the automated power-law fit procedure, which has already been described in section 2.

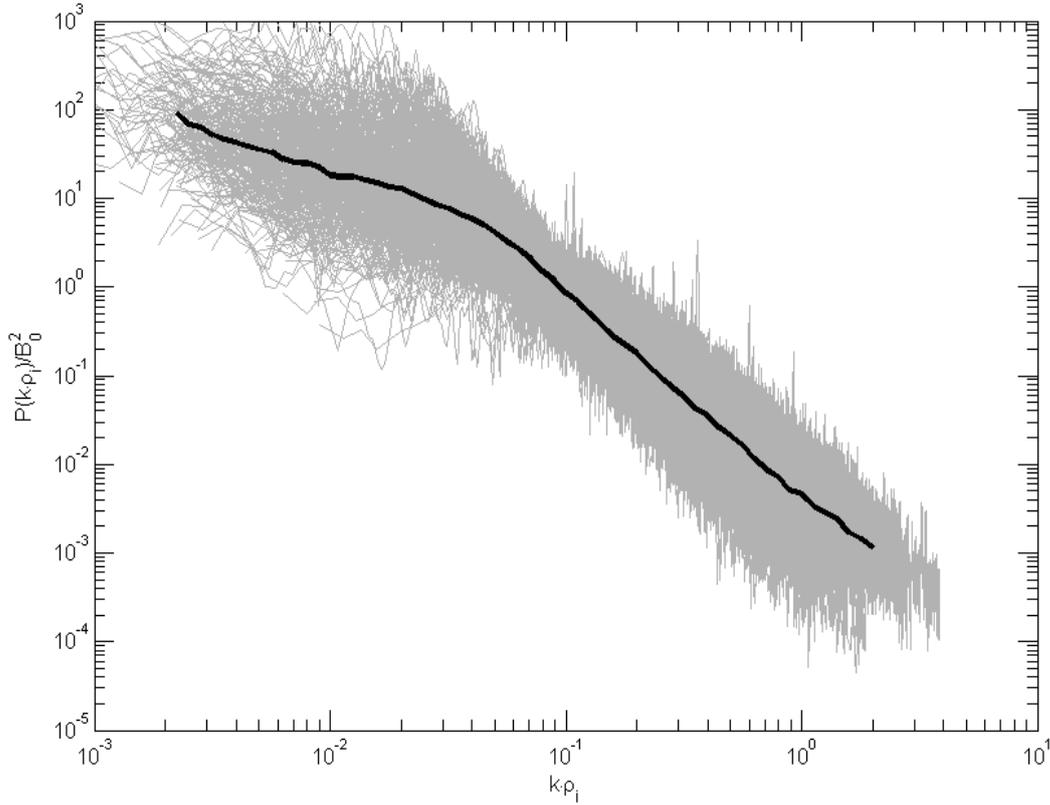

Figure 2. *Normalised superposed power spectral density of the magnetic field time-series (337 events) inside the magnetosheath. The spatial coordinate spanning the streamline component of the wave number, $k = \frac{2\pi f_{sc}}{V_{flow}}$, in the unit of ion gyroradius $\rho_i = \frac{V_{T_i}}{2\pi f_{ci}}$ is depicted on the x-axis, and the magnetic field power spectral density, normalised to $B_0^2$, is shown on the y-axis.*

Figure 3 represents the 2D colour-coded histograms for the variation of the spectral slope values ($\alpha$) of the magnetic field spectra in the magnetosheath region of Earth. The frequency, $f_{sc}$, is normalised to the ion gyrofrequency, $f_{ci}$, and is depicted on the x-axis, the spectral slope distribution is shown on the y-axis, and the number of occurrences of different spectral slope values in logarithmically equal x-axis bins is represented by a colour bar. We observe the power-law at low frequencies, $f_{sc}/f_{ci} < 0.5$, with a uniform distribution of spectral index varying between -1.5 and 0. There is a sudden modification in $\alpha$ from low, $f_{sc}/f_{ci} < 0.5$, to high frequencies, $f_{sc}/f_{ci} > 0.5$, with spectral slope value ranging from about -2.6 to -1.8. A persistent power-law is detected in the broad frequency range after $f_{sc}/f_{ci} = 0.5$, and the value of $\alpha$ is close to -2.4 for the most dominant peak.

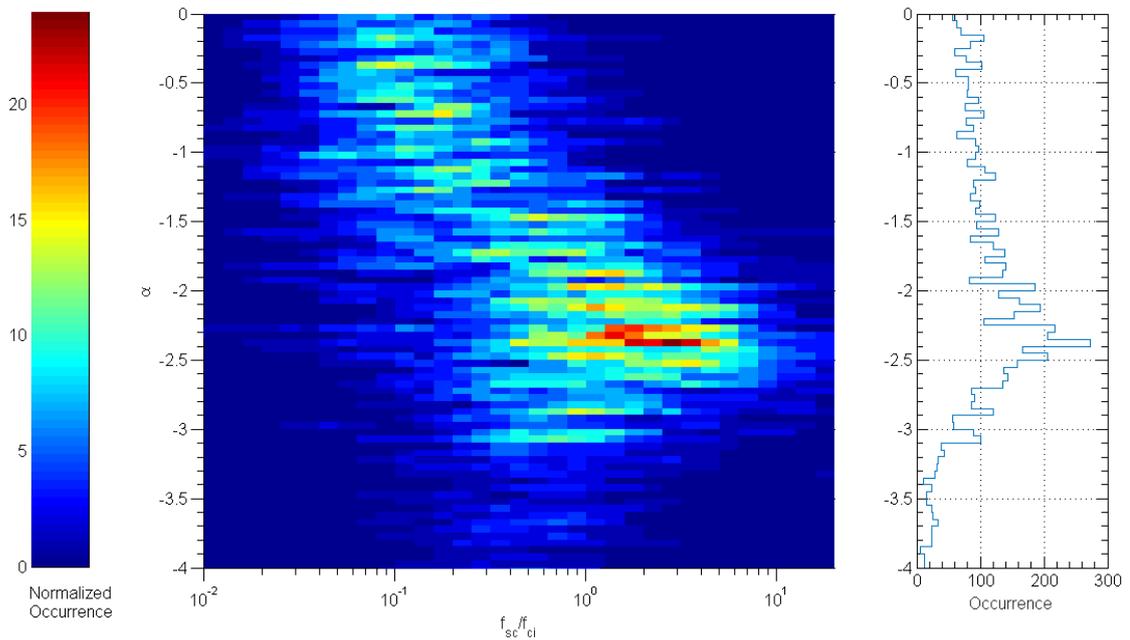

Figure 3. *Two-dimensional histogram for the spectral slope analysis of the magnetic field fluctuations, frequency, $f_{sc}$, normalised to ion gyrofrequency, $f_{ci}$, is presented on the x-axis, different slope values are depicted on the y-axis, and colours show the cumulative number of occurrence of different slopes at logarithmically equal frequency bins. The right histogram shows the cumulative number of occurrence of different slope values in the given frequency range.*

Additionally, the 2D histogram of spectral slope values obtained in the wavenumber domain for the Earth's magnetosheath is shown in Fig. 4, in which the streamline wave number, $k$, calculated in the unit of $\rho_i$ is on the *x*-axis, $\alpha$ is on the *y*-axis, and colours show the cumulative number of slope occurrences in logarithmically equal wave number bins. The spectral break is found at around $k\rho_i = 0.05$, accompanied by steepened spectra with about $\alpha = -2.6$ to $-1.6$ at smaller scales. However, at larger scales $k\rho_i < 0.05$, $\alpha$ exhibits a nearly uniform distribution between -1.5 and 0.

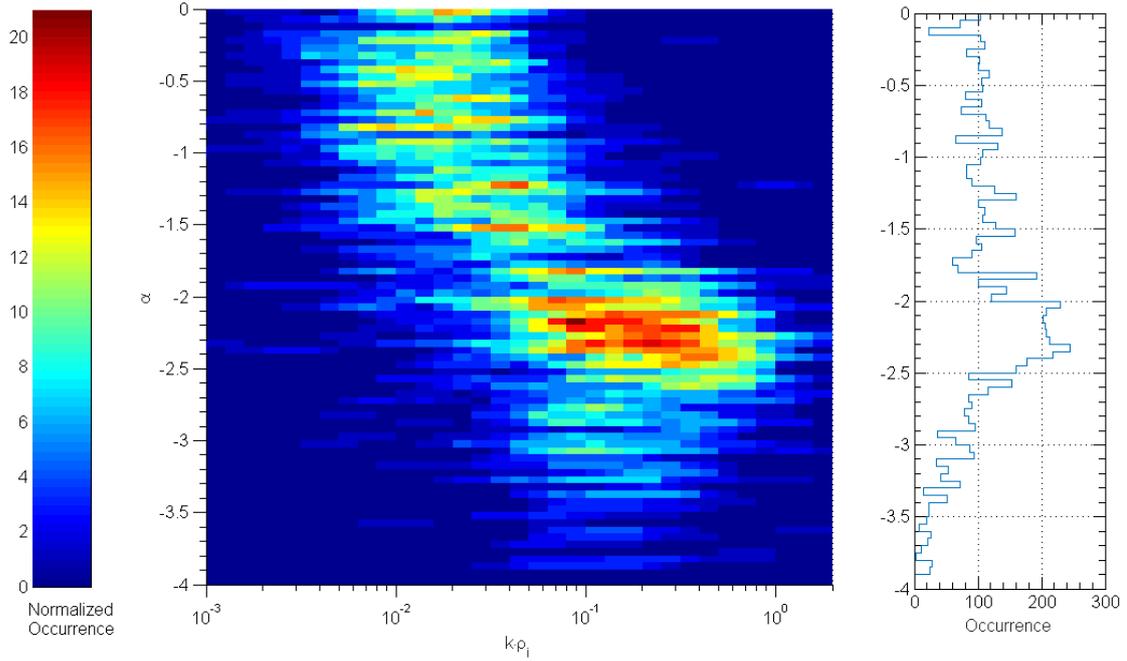

Figure 4. *Two-dimensional histogram for the spectral slope analysis of the magnetic field fluctuations is shown in the spatial spanning of wave number, $k$, in the unit of ion gyroradius, $\rho_i$, and the cumulative number of occurrence of different slopes at logarithmically equal wave number bins is shown by a colour bar. The right histogram shows the cumulative number of occurrence of different slope values in the given wave number.*

Figures 5-6, show the normalised numbers of occurrences of the different propagation angle, $\theta_{kB}$, and compressibility values, $R_\parallel$, as a function of frequency, $f_{sc}/f_{ci}$. The values of each pixel are normalised by the total number of events observed in a frequency bin. Figure 5 exemplifies that at low frequencies, $f_{sc}/f_{ci} < 0.3$, the most frequent values of $\theta_{kB}$ are close to 90° suggesting wave vectors predominantly perpendicular to $B_0$. For frequencies between $0.2 f_{ci}$

and $2f_{ci}$, the waves/fluctuations are organised in two populations either nearly parallel or perpendicular. For frequencies between $2f_{ci}$ and $10f_{ci}$, the distribution of $\theta_{kB}$ spans a narrower range (between $75°$ and $90°$) and has a peak for perpendicular direction.

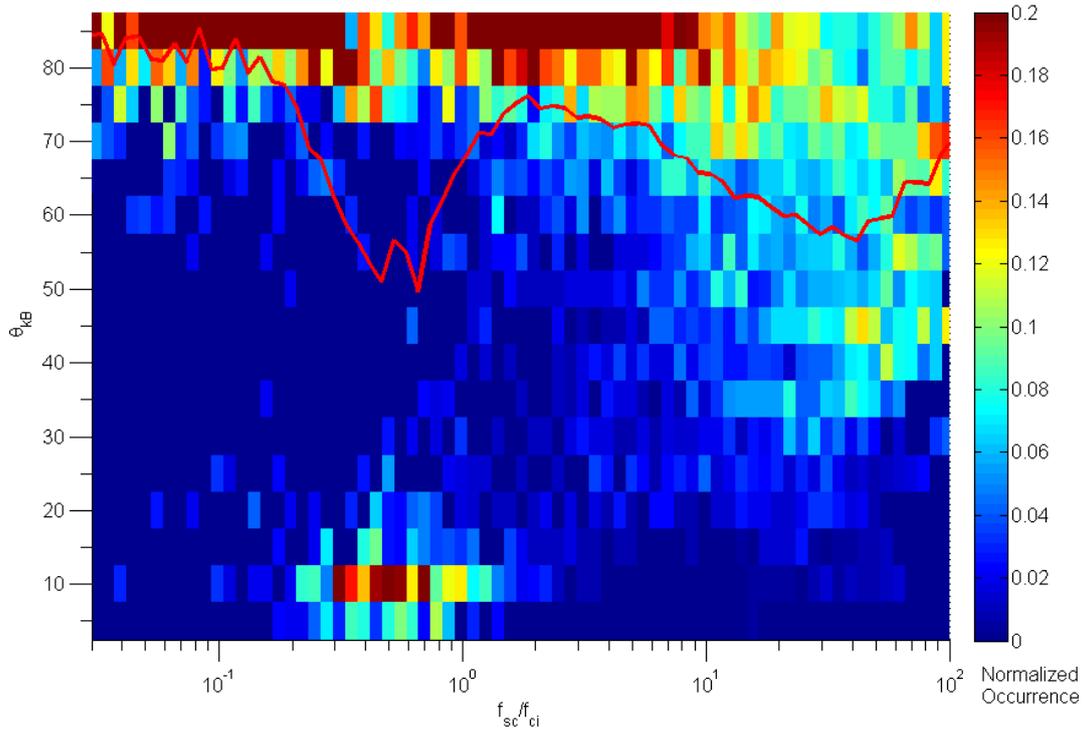

Figure 5. *Two-dimensional histogram of the distribution of the angle of propagation ($\theta_{kB}$) concerning frequency ($f_{sc}/f_{ci}$).*

For frequencies above $10f_{ci}$, the distribution becomes broader and the peak at $90°$ vanishes gradually at higher frequencies. The thick red line shows the weighted variation of $\theta_{kB}$ in given frequency range.

Figure 6 shows the statistical distribution of compressibility as a function of frequency. The colours represent normalised occurrences of compressibility values. At low frequencies below $f_{sc}/f_{ci}=0.2$, the compressibility has a broad range of variation ranging from 0.1 to 0.9. At frequencies where parallel propagation is observed, i.e., $0.2 < f_{sc}/f_{ci} < 2$ (see Fig. 5), a distinct population of waves appears with very low compressibility.

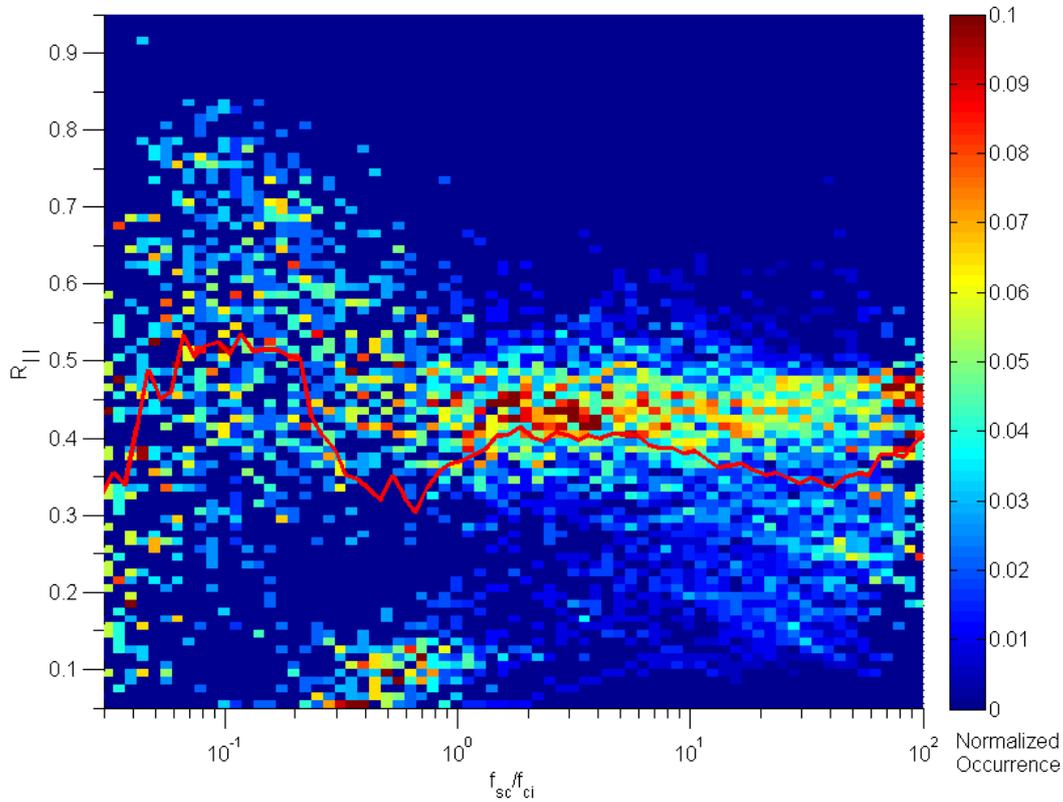

Figure 6. *Two-dimensional histogram of the distribution of compressibility ( $R_{\parallel}$ ) concerning frequency ( $f_{sc}/f_{ci}$ ).*

Ion cyclotron mode is incompressible, circularly or elliptically (left-hand) polarised, and the direction of the wave vectors can change from parallel to quasi-perpendicular (up to 70° or 80°, above which the mode becomes kinetic Alfvén mode). These two findings imply the existence of transverse Alfvèn waves (Zaqarashvili et al. 2011, 2013, Sharma & Singh 2008, Dwivedi et al. 2012, 2013, Sharma & Modi 2013) near the proton cyclotron frequency. For frequencies higher than the frequency $f_{sc}/f_{ci} = 2$, $R_{\parallel}$ takes the values between 0.2 and 0.5. Similar to Fig.5, the thick red line illustrates the weighted variation in $R_{\parallel}$ over the given frequency band.

## 4.2 Numerical simulation results

Here we present the simulation result of KAW's magnetic field spectra in the intermediate-$\beta$ regime applicable to magnetosheath region of the Earth. Figure 7 depicts the variation of $|B_{yk}|^2$ against parallel wave number $k_z$ (represented $k$ in Fig. 7) at $k_x = 0$. It is clear from the figure that the magnetic field spectra have the spectral slope $\approx -2.8$, that is close to the spectral slope range -2.6 to -1.8 as observed from the Cluster data at the small scale (see Fig. 4).

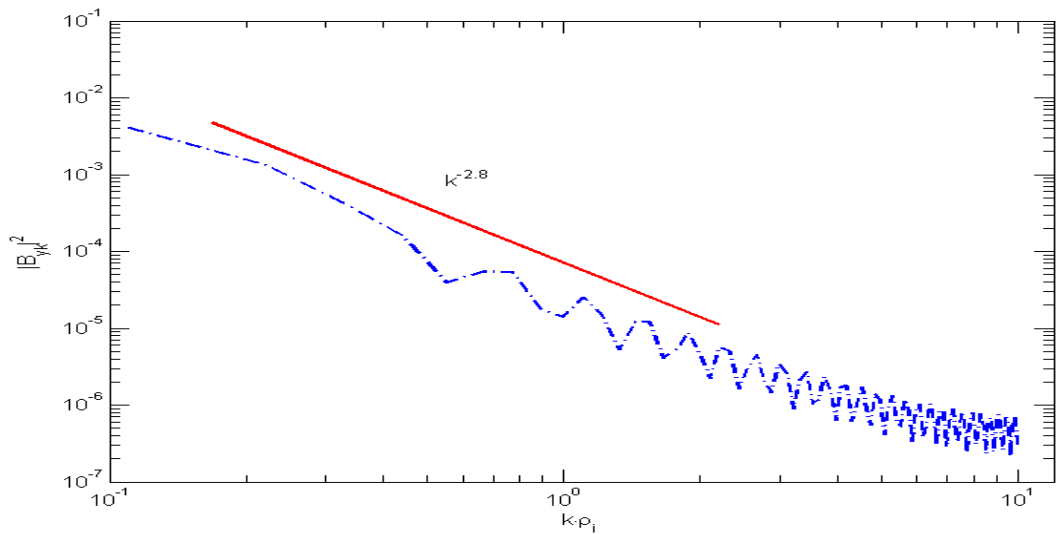

Figure 7. *The magnetic field spectra obtained by the simulation. The variation of $|B_{yk}|^2$ against $k$ is depicted on the y-axis. The thick red line represents the spectral slope value.*

Unlike many of other simulation studies (Karimabadi et al. 2013; Boldyrev et al. 2013; Franci et al. 2018; Groseli et al. 2018) that have looked at KAW spectra, the model and numerical approach discussed in the present paper is unique in the sense that it is based on two-fluid description which is fairly valid up to ion scale and therefore quite efficiently describe the wave and turbulence processes up to ion scale without any significant enhancement on computing power.

## 5. Discussion

Cluster observations of the magnetic field fluctuations reveal in a majority of cases that for frequencies smaller than $0.5 f_{ci}$, the spectral slope takes values mostly between -1.5 and 0. The distribution of spectral slopes in this range is rather flat and show no peak. For frequencies higher than $0.5 f_{ci}$, the spectra exhibit power-law behaviour, and the distribution of the spectral slopes in this range show a peak at $\alpha = -2.4$. The transition between the two regimes is rapid and takes place for frequencies around $0.5 f_{ci}$. In the spacecraft moving frame of reference, the spectral break is detected at around $k\rho_i = 0.05$. In the real solar wind magnetic field time-series, the spectral break in the magnetic field spectra is observed close to ion gyroradius (Chen et al., 2014) when plasma $\beta$ is greater than one. However in case of plasma $\beta < 1$, the spectral break is detected in the vicinity of ion inertial length. Franci et al. (2016) further realise the similar results with the particle-in-cell simulations. In the present work, we have selected 337 magnetosheath events, which provide us a unique opportunity to revisit the abovementioned facts. However, while re-examining the selected magnetosheath events based on the plasma $\beta$ we found that only in 13 events out of 337 events the plasma $\beta$ was less than 1. The median of $\beta$ values for all time intervals is found to be 4.21, and we define this value as the lower limit for the selection of high plasma $\beta$ time-series. In a supplementary analysis we further analyse the statistical distribution of the spectral slopes in terms of spatial scales, for the time-series belonging to plasma $\beta > 4.21$. We found that neither the average length scale of the spectral breaks nor the representative spectral slopes at large and small scales (the results are not given in the paper) exhibit any significant change in comparison with the results obtained from the statistical analysis of the whole set (337) of intervals.

In addition to this, a conditional analysis is performed to check the statistical reliability of the Taylor hypothesis in the case of our set of magnetosheath time-series. Therefore, the slope statistics (scaling ranges and spectral exponents shown in Figs. 3 and 4) are further investigated for the magentosheath time intervals having $V_{flow} < 150\ km/s$ (111 events) and $V_{flow} > 219\ km/s$ (112 events). We found (the results are not shown in the paper) that the basic statistical results (typical scaling ranges, place of spectral break and typical slope values) remain unchanged for the sets of events with different bulk flow velocities. Thus, we conclude that in most of the investigated intervals, the plasma flow is fast enough for evaluating the wave properties in spatial scales from the spacecraft observations.

One of the most important findings of the statistical study of our magnetosheath time-series is that we do not retrieve a significant occurrence of the Kolmogorov scaling as we observed in the solar wind spectra. The steepening in the energy spectra could be attributed to kinetic Alfvèn wave turbulence (Leamon et al., 1999; Sharma & Singh 2008; Dwivedi et al. 2012, 2013; Sharma & Modi 2013; Dwivedi et al., 2013). The absence of Kolmogorov scaling in most of the cases may question the very existence of the turbulent nonlinear cascade in the Earth's magnetosheath. One possible reason for not observing the inertial turbulent Cascade Range in the terrestrial magnetosheath spectra could be associated with the dynamics dominated by the energy of the wave activities (either ion cyclotron or mirror mode) over the low amplitude turbulent fluctuations (Kovacs et al. 2014). Another reason for the lack of a Kolmogorov scaling in the magnetosheath may be related to the amount of time the large scales have had to nonlinearly interact since the plasma passed through the bow shock (Hadid et al. 2015; Huang et al. 2017; Chhiber 2018). That is to say that the large-scales haven't had time to fully develop since the plasma was 'stirred up' by the shock.

Indeed at low frequencies, $f < 0.2 f_{ci}$, the distribution of $\theta_{kB}$ has a maximum near to 90°. The compressibility factor is distributed between 0.1 and 0.9. Thus the fluctuations show no prevalence of compressive or transversal fluctuations. In the frequency range of $0.2 \leq f_{sc}/f_{ci} \leq 2$, the histograms of both, the propagation angle (Fig. 5) and compressibility (Fig. 6) show the occurrence of two different populations of wave modes. We suppose that the first population exhibits perpendicular propagation and intermediate compressibility, while the second population contains transversal waves with small propagation angle. The transversal and parallel propagating wave mode below the ion cyclotron frequency can suggest the turbulence is Alfvènic. The propagation of waves between $2f_{ci}$ and $10f_{ci}$, is mostly perpendicular to the ambient magnetic field, while at frequencies higher than $10f_{ci}$ the waves are oblique to $B_0$ with a wide distribution of propagation angles. The compressibility, on the other hand, shows rather uniform values in the range of about $0.2 \leq R_\parallel \leq 0.5$ which implies the isotropic distribution of wave energies. The simulation result reveals that KAW magnetic field spectrum has the spectral index of $k^{-2.8}$ that is close to the observed spectral indices. We argue that at high frequencies, perpendicular Alfvèn waves (Sharma & Singh 2008) could contribute to the observed statistical fluctuations properties.

## 6. Conclusion

In the present work, we investigate the turbulent dynamics of the magnetosheath plasma using spectral analysis of magnetic field fluctuations. Specifically, we analyse the relevant scales associated with changes in the spectral behaviour. The persistent detection of power-laws in the power spectra of magnetic field fluctuations and different scaling laws for the different range of frequencies/scales has been found. The statistical analysis of the spectra of 337 time-series evidences the existence of a spectral range with the most probable spectral exponent close to -2.4 scaling of fully developed inertial range turbulence. The scaling range extends for about one decade both in the frequency and wavenumber domains at scales around the ion cyclotron frequency and ion gyroradius length. The theoretical model and simulated magnetic field spectra show that the nonlinear KAW can describe the nature of the small-scale turbulence.

*Acknowledgements*

*The authors would like to thank the entire members of Cluster fluxgate magnetometer team, and Cluster Science Archive for providing the data used in the present work. The magnetosheath data set was compiled in the frame of the STORM project (grant agreement No. 313038). The Austrian Science Foundation (FWF) (Project I2939-N27), Austrian Agency for International Cooperation in Education and Research Project No. IN 05/2018, and Austrian Academy of Sciences have supported this work. MLK acknowledges the support from the project S11606-N16 and partial support by the Ministry of Education and Science of Russian Federation (Grant No. RFMEFI61617X0084).*